\documentstyle[12pt]{article}

\input{tcilatex}

\begin{document}

\title{Static Generalized Brans-Dicke Universe and Gravitational Waves Amplification}
\author{Marcelo.S.Berman$^{(1)}$ \and and Luis.A.Trevisan$^{(2)}$ \\
(1) Instituto de Tecnologia do Paran\'{a}-\\
-Tecpar-Grupo de Projetos Especiais.\\
R.Prof Algacyr M. Mader 3775--CIC-CEP 81350-010\\
Curitiba-PR-Brasil\\
Email: marsambe@tecpar.br\\
(2) Universidade Estadual de Ponta Grossa,\\
Demat, CEP 84010-330, Ponta Grossa,Pr,\\
Brasil \ email: latrevis@uepg.br}
\maketitle

\begin{abstract}
We \ present a Static Universe in a generalized Brans-Dicke gravity theory,
where the coupling ``constant \ '' varies with time, as well as the scalar
field. There \ is amplificatiom of gravitational waves in such Universe, at
least when it is ``young''.

\end{abstract}

\newpage

\section{Introduction}

\noindent\ The first models of the Universe that we can find in the
literature are Einstein's static ones (see, for instance, Tolman \cite{1}).

Today, the study of static models has a two-fold aim. First, it throws light
on the theory, and one can compare with expansion models and conclude which
physical effects are due on the expanding properties of scale-factors, and
which are not. Second, we would like to remind that, according to some
critical approaches, it is to be questioned, at least just for its own sake,
whether there could be an explanation to ``cosmological '' red-shifts, other
than a recessional effect.

With these objectives in mind, Berman, Som and Gomide studied static
Brans-Dicke models\cite{2}; Berman \cite{3} studied their stability
properties, too. Berman also studied \cite{4} such Universes in ``modified
'' \ B.D. theory, and a static Universe with a magnetic field in Einstein
Cartan's Cosmology \cite{5}. \ Afterwards, Berman and Som \cite{6} also
delved into the static solutions in Wesson's 5D theory of gravity. Earlier,
Berman \cite{7} also studied inhomogeneous static models in B.D theory.

All these studies reveal details of theoretical characteristics not
necessarily associated with evolutionary models. We shall now study a static
model with time varying coupling ''constant'', as well as a variable scalar
field. Scalar tensor cosmologies were earlier studied by Barrow \cite{8}. We
wish to point out that the core of the theory on scalar-tensor cosmologies
was undertaken by J.D.Barrow and collaborators(see for instance references 
\cite{9}\cite{11}).

\section{The Field Equations}

Let us consider the following action:

\begin{equation}
L_{\Phi }=-\Phi R+\Phi ^{-1}w(\Phi )\partial _{a}{\Phi }\partial ^{a}{\Phi }%
+16{\pi }L_{m}
\end{equation}
where $L_{m}$ is the Lagrangian for matter fields, and $\Phi $ is the scalar
field. If $w=const$ we obtain the Brans-Dicke theory \cite{10}. This
Lagrangian was adopted by Barrow and Maeda \cite{9} . For a discussion about
the Lagrangians of the scalar theories of gravitation, see \cite{12}.

We now shall undertake the study of such Universes in an $\omega =\omega
(\phi )$ Brans-Dicke generalized theory, the field equations being:

\begin{equation}
G_{ab}=-\frac{8\pi }{\phi }T_{ab}-\frac{\omega }{\phi ^{2}}\left[ \phi
_{a}\phi _{b}-\frac{1}{2}g_{ab}\phi _{i}\phi ^{i}\right] -\frac{1}{\phi }%
\left[ \phi _{a;b}-g_{ab}\Box \phi \right]
\end{equation}
and

\begin{equation}
\left[ 3+2\omega \right] \Box \phi =8\pi T-\left( \frac{d\omega }{d\phi }%
\right) \phi _{i}\phi ^{i}
\end{equation}
while we must impose conservation of energy-momentum tensor:

\begin{equation}
T_{;b}^{ab}=0
\end{equation}

For \ a static metric, and a flat Universe,

\begin{equation}
ds^{2}=dt^{2}-R_{0}^{2}\left[ dx^{2}+dy^{2}+dz^{2}\right]
\end{equation}
we find, then:

\begin{equation}
\frac{8\pi \rho _{0}}{\phi }+\frac{\omega }{6}\frac{\dot{\phi}}{\phi^{2}}=0
\end{equation}

\begin{equation}
\ddot{\phi}=-\frac{\dot{\phi}\dot\omega}{3+2\omega }+\frac{8\pi }{3+2\omega }%
\left( \rho _{0}-3p_{0}\right)
\end{equation}
where $\rho =\rho _{0}=const.$

For a perfect gas equation of state

\begin{equation}
p=\alpha \rho =p_{0}=const.
\end{equation}

Let us suppose that:$\ddot{\phi}=0$ and $\omega \neq -3/2.$ We find the
solution :

\begin{equation}
\omega =8\pi \left( \rho _{0}-3p_{0}\right) t
\end{equation}
and

\begin{equation}
{\phi} =\frac{6\rho _{0}}{3p_{0}-\rho _{0}}t
\end{equation}

The solution is time-dependent, although the scale-factor is constant, like
the energy density and pressure. If $\rho _{0}>3p_{0,\text{ \ }}$we have $%
\omega \geq 0,$ and:

\begin{equation}
\lim_{t\rightarrow \infty }\omega =\infty ,
\end{equation}
which means that General \ Relativity is the limiting gravity theory, for an
``old '' Universe.

It is clear that the time variation of $\omega $ and $\phi $ cannot be
attributed to an evolution in the scale-factor.

\section{Gravitational Waves.}

Barrow et al. \cite{11} calculated \ the amplification equation for
gravitational waves, which has here the following form:

\begin{equation}
\ddot{Y_{k}}+\frac{\dot{\phi}}{\phi }\dot{Y_{k}}+\left[ \frac{k^{2}}{%
R_{0}^{2}}-\frac{\dot{\omega}}{2\omega +3}\frac{\dot{\phi}}{\phi }\right]
Y_{k}=0
\end{equation}
where $k=|\overrightarrow{k}|=2\pi R_{0}/\lambda .$

For an ``old '' Universe, we would have a non-amplified harmonic function,
while for a young Universe, we are left with the equation:

\begin{equation}
\ddot{Y_{k}}+t^{-1}\dot{Y_{k}}-\frac{1}{2}t^{-2}Y_{k}\cong 0
\end{equation}
which has the solution

\begin{equation}
Y_{k}=At^{\frac{\sqrt{2}}{2}}
\end{equation}
($A=const),$ and then we have amplification of gravitational waves.

\section{Conclusion}

We studied a Static generalized Brans-Dicke theory. It is shown that the
time variation of the scalar field $\phi $ and of the coupling ``constant '' 
$\omega $ are not a consequence of the variation of the scale-factor, which
is constant, in this work, by hypothesis. Another effect of such model is
the amplification of gravitational \ waves in the early Universe.

\textbf{Acknowledgments}

Both authors thank support by Prof Ramiro Wahrhftifig, Secretary of Science
,Technology and Higher Education of the State of Parana, and by our
Institutions, especially to Jorge\ L.Valgas, Roberto Merhy, Mauro K.
Nagashima, Carlos Fior, C.R. Kloss, J.L.Buso, and Roberto Almeida.

\end{document}